\documentclass[a4paper]{spie}  

\usepackage[]{graphicx}
\usepackage{units}
\usepackage{fancyvrb}
\usepackage{mathtools}
\usepackage{amsmath}
\usepackage{amssymb}
\newbox\verbbox
\title{Shack-Hartmann wavefront sensor sensitivity loss factor estimation in partial correction regime}

\author[a,c]{Guido Agapito}
\author[b,c]{Carmelo Arcidiacono}
\author[a,c]{Simone Esposito}

\affil[a]{INAF -- Osservatorio Astrofisico di Arcetri, Largo E. Fermi 5, 50125 Firenze, Italy}
\affil[b]{INAF -- Osservatorio Astronomico di Bologna, Via P. Gobetti 93/3, 40129 Bologna, Italy}
\affil[c]{ADONI -- Laboratorio Nazionale di Ottica Adattiva, Italy}

\authorinfo{Further author information: (Send correspondence to G. Agapito)\\G. Agapito: E-mail: agapito@arcetri.astro.it}

\begin{document}

\def\arcsec{$^{\prime\prime}$}
\def\araa{Annual Review of Astronomy and Astrophysics}
\def\apj{The Astrophysical Journal}

\maketitle 

\begin{abstract}
In typical adaptive optics applications, the atmospheric residual turbulence affects the wavefront sensor response decreasing its sensitivity. On the other hand, wavefront sensors are generally calibrated in diffraction limited condition, and, so, the interaction matrix sensitivity may differ from the closed loop one. The ratio between the two sensitivities, that we will call the sensitivity loss factor, has to be estimated to retrieve a well-calibrated measurement. The spots size measurement could give a good estimation, but it is limited to systems with spots well sampled and uniform across the pupil. We present an algorithm to estimate the sensitivity loss factor from closed loop data, based on the known parameters of the closed loop transfer functions. Here we preferred for simplicity the Shack-Hartmann WFS, but the algorithm we propose can be extended to other WFSs. 

\end{abstract}

\section{Introduction}

In this article we focus on the sensitivity gain of the Wavefront Sensor (WFS),
that is the ratio between the calibrated sensitivity and
the one obtained during operation.
For many applications we need the actual WFS sensitivity
to get the correct measurement (very important in open-loop),
to proper optimize the temporal controller,
and to proper compensate for Non Common Path Aberrations (NCPA).   
A change on the WFS sensitivity can be caused by a variation of 
the spot size\cite{2008MNRAS.387..173T} (strong effect in a quadcell SH) or a change of dark
current and/or sky background.
In some cases these effects can be measured and so the sensitivity
loss factor can be estimated \cite{Fugate:94, doi:10.1117/12.217763, doi:10.1117/12.321745}, otherwise a specific algorithm
should be used to estimate this \cite{Veran:00, EspositoNCPA2016}. 

In this paper we present an approach to the estimation of this coefficient
based on the closed loop Transfer Function (TF) and on the loop data Power Spectral Densities (PSD).


\section{Closed Loop Transfer Functions and WFS sensitivity}\label{CLTF}
\begin{figure*}
	\begin{center}
		\includegraphics[width=0.85\columnwidth]{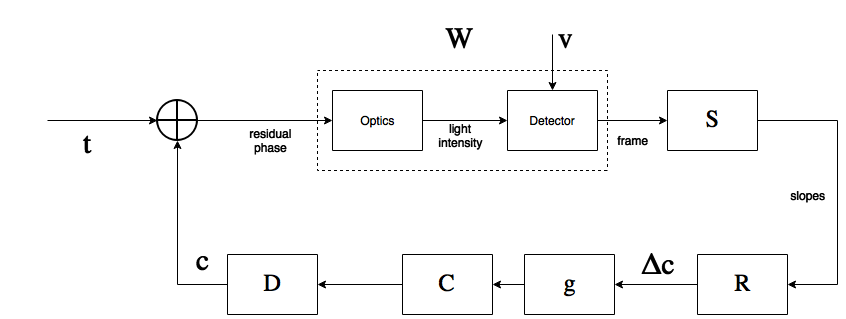}
		\caption{Scheme of a SCAO closed loop system.}
		\label{fig:SCAO}
	\end{center}
\end{figure*}
Let us consider a Single Conjugate Adaptive Optics (SCAO) closed loop system
as the one shown in Figure \ref{fig:SCAO}.
Note that we choose a SCAO system for simplicity, however other systems,
like multi-conjugate adaptive optics\cite{beckers88,beckers89a} MCAO, can also be considered as well.
The command vector $c$, and the incremental command vector $\Delta c$ are defined as:
\begin{equation}\label{eq:C}
    c = \frac{gH}{1 + g H} t + \frac{gS R C D}{1 + g H} v \, ,
\end{equation}
\begin{equation}\label{eq:DeltaC}
    \Delta c = \frac{W S R}{1 + g H} t + \frac{S R}{1 + g H} v \, ,
\end{equation}
where:
\begin{equation}\label{eq:H}
    H = W S R C D \, ,
\end{equation} 
$W$ is the WFS (optics + detector), $S$ is the slope computer,
$R$ is the reconstruction matrix, $D$ is the deformable mirror,
$g$ is the integrator gain vector, $C$ is the controller,
$t$ is the turbulence and $v$ the measurement noise vector.\\
Note that the Rejection Transfer Function (RTF)
$H_{R}=\frac{1}{1 + g H}$, and the Noise Transfer Function (NTF)
$H_{N}=\frac{g H}{1 + g H}$ are implicitly present in Equations
\ref{eq:C} and \ref{eq:DeltaC}.

Then, we add an unknown linear coefficient in the WFS Transfer Function (TF),
$\alpha$, that will be called sensitivity
loss factor.
Let us define $W=\alpha W'$, so the previous equations become:
\begin{equation}\label{eq:Halpha}
    H =\alpha W' S R C D = \alpha H' \, ,
\end{equation}
\begin{equation}\label{eq:Calpha}
    c = \frac{g\alpha H'}{1 + g \alpha H'} t + \frac{gS R C D}{1 + g \alpha H'} v \, .
\end{equation}
\begin{equation}\label{eq:DeltaCalpha}
    \Delta c = \frac{\alpha W' S R}{1 + g \alpha H'} t + \frac{S R}{1 + g \alpha H'} v \, ,
\end{equation}

\begin{figure*}
    \begin{center}
        \begin{minipage}[b]{.46\textwidth}
            \includegraphics[width=\textwidth]{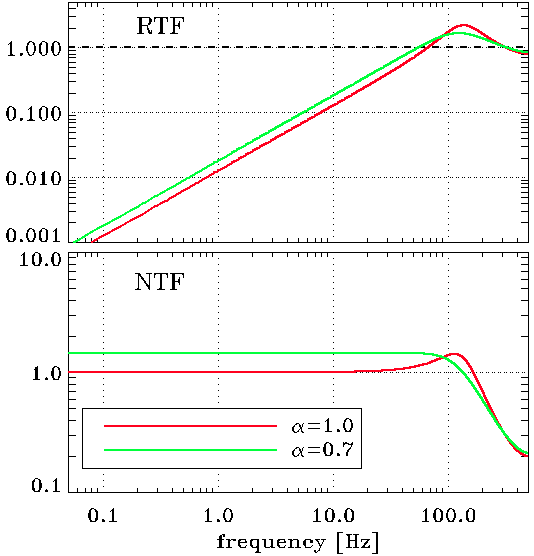}
        \end{minipage}
        \begin{minipage}[b]{.53\textwidth}
            \includegraphics[width=\textwidth]{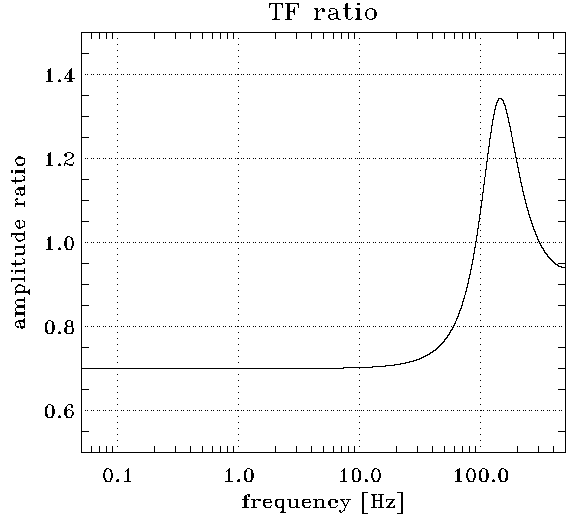}
        \end{minipage}
        \caption{Left: RTFs and NTFs of an AO system with a sampling frequency of 1000Hz,
        2 frames of delay, an integrator gain of 0.5 and different WFS sensitivity loss
        factors (1 and 0.7).
        Right: Ratio between RTFs (or NTFs) shown on the left.}
        \label{fig:TFwfsSens}
    \end{center}
\end{figure*}
\begin{figure*}
	\begin{center}
		\begin{minipage}[b]{.49\textwidth}
			\includegraphics[width=\textwidth]{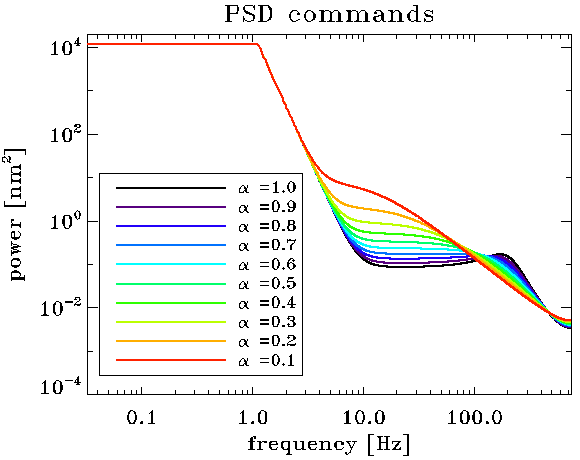}
		\end{minipage}
		\begin{minipage}[b]{.49\textwidth}
			\includegraphics[width=\textwidth]{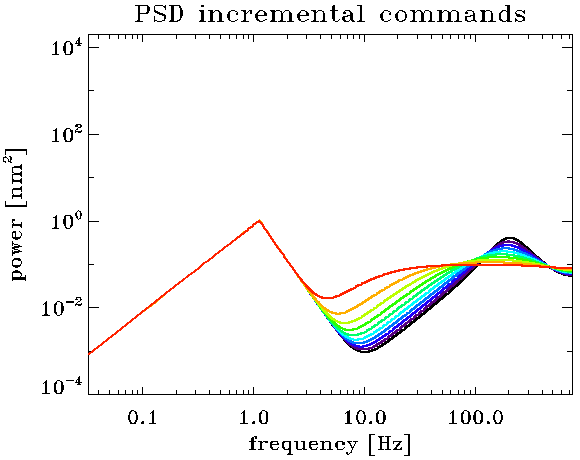}
		\end{minipage}
		\caption{Effects of the sensitivity loss factor $\alpha$. Left: Effect on Tilt command.
		Right: Effect on Tilt incremental command.
		Turbulence $r_0$ is 0.15m, measurement noise variance is 60nm${}^2$, integrator gain is 0.7 and total delay is 2 frames.}
		\label{fig:sensLossEff}
	\end{center}
\end{figure*}

We set $S=1$, $R=1$, that means they have no dynamics,
$W'=D=z^{-1}$, that is a pure one step delay dynamics
(for a total of two steps of delay), and $C=\frac{1}{1-z^{-1}}$,
that is a pure integrator, equations \ref{eq:DeltaCalpha} and \ref{eq:Calpha} become:
%
%
\begin{equation}\label{eq:Calpha2frame+}
    c = \frac{g\alpha}{z^{2} - z + g \alpha} t + \frac{g z}{z^{2} - z + g \alpha} v \, .
\end{equation}
\begin{equation}\label{eq:DeltaCalpha2frame+}
    \Delta c = \frac{\alpha (z-1)}{z^{2} - z + g \alpha} t + \frac{z(z-1)}{z^{2} - z + g \alpha} v \,
\end{equation}
%
%
%
%
%
%
%
%
%
%
%
%
The effect of the sensitivity loss factor on RTF and NTF is shown in Figure~\ref{fig:TFwfsSens}, and on command and incremental command in
Figure~\ref{fig:sensLossEff}.\\

Note that it is possible to estimate the RTF, as show in Dessenne et al. \cite{1998ApOpt..37.4623D}, but, unfortunately, this estimation is not affected by the sensitivity loss factor:
in fact, if we compute this TF using Equations~\ref{eq:Calpha} and~\ref{eq:DeltaCalpha},
and $z^{-k}=W'$, that is the delay to be added to the commands to be synchronized with
the incremental commands, we get:
\begin{equation} \label{eq:TFestimated}
	\frac{\Delta c}{\Delta c + c z^{-k}} = \frac{1}{1+gCDz^{-k}} \, ,
\end{equation}
and so, $\alpha$ has been reduced in the computation.

\section{Sensitivity estimation}

The sensitivity loss factor, $\alpha$, estimation is evaluated minimizing the cost function in the
frequency range $\left[\omega_1,\omega_2\right]$:
\begin{equation} \label{eq:cost}
	J(\alpha) = \sum_{\omega=\omega_1}^{\omega_2}{( \mathrm{PSD}\{ \Delta c(\omega) \} -
	\hat{K}(\omega,\alpha))^2} \, ,
\end{equation}
where:
\begin{equation}\label{eq:Komega}
	\begin{multlined}
		\hat{K}(\omega,\alpha) = \left( \frac{\alpha W' S R}{1 + g \alpha H'} \right)^{2} \mathrm{PSD}\{ t(\omega)\} + \\
		\left( \frac{S R}{1 + g \alpha H'} \right)^{2}  \mathrm{PSD}\{ v(\omega)\} \, ,
	\end{multlined}
\end{equation}
and the operator:
\begin{equation} \label{eq:PSD}
    \mathrm{PSD}\{ x(\omega) \} \triangleq \langle | x(\omega) | ^2  \rangle
\end{equation}
denotes the temporal Power Spectral Density (PSD) of $x$.\\
So, summarizing, we search for the sensitivity loss factor, $\alpha$, which minimizes the difference between
the incremental command PSD and the PSD, computed filtering the inputs PSDs by the $\alpha$ dependent
TFs.\\
Note that here we suppose to know exactly both the turbulence $t$ and the measurement noise $v$.
In a real system these data should be estimated before running the sensitivity loss factor
estimation.\\

\section{Examples}
\begin{figure*}
	\begin{center}
		\begin{minipage}[b]{.49\textwidth}
			\includegraphics[width=\textwidth]{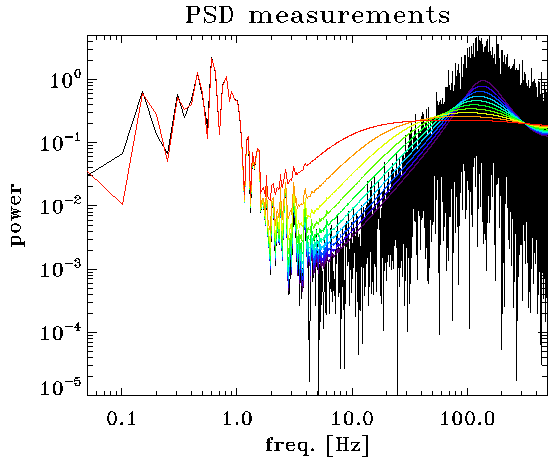}
		\end{minipage}
		\begin{minipage}[b]{.49\textwidth}
			\includegraphics[width=\textwidth]{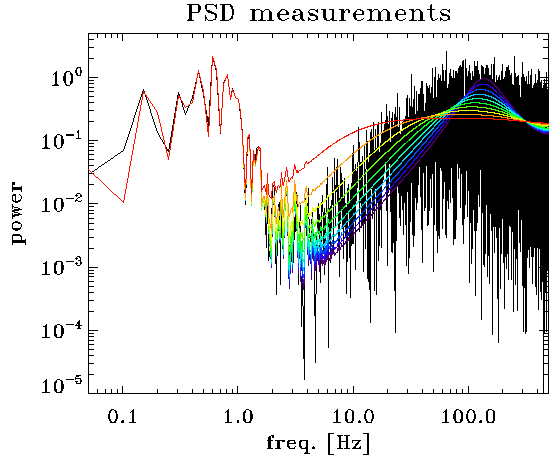}
		\end{minipage}
		\caption{Effects of a spot size on the closed loop measurement PSD.
		Left: spot size as in calibration.
		Right: bigger spot size (spot convoled with gaussian with a FWHM of 8 pixels, corresponding to 1,84arcsec).
		Input Tip-Tilt RMS is 255nm, measurement noise variance is 10nm${}^2$, integrator gain is 0.5 and
		total delay is 2 frames. $\alpha=1$ means no sensitivity gain change.}
		\label{fig:quadCellSpot}
	\end{center}
\end{figure*}
\begin{figure*}
	\begin{center}
		\begin{minipage}[b]{.49\textwidth}
			\includegraphics[width=\textwidth]{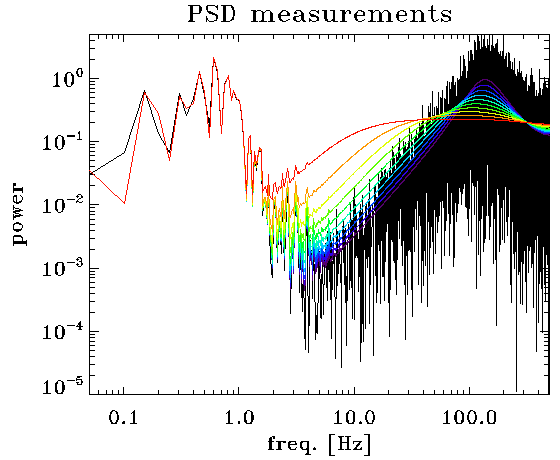}
		\end{minipage}
		\begin{minipage}[b]{.49\textwidth}
			\includegraphics[width=\textwidth]{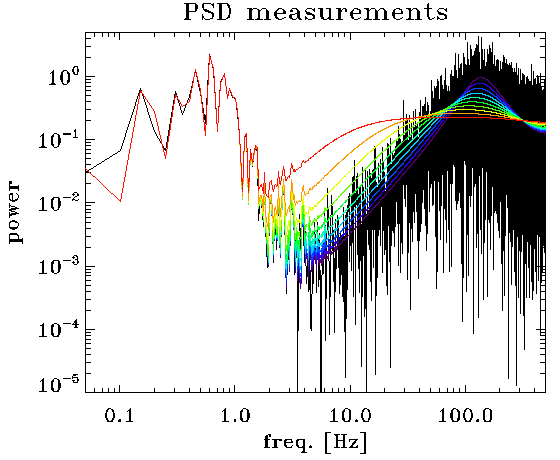}
		\end{minipage}
		\caption{Effects of a strong background on the closed loop measurement PSD.
		Left: no background as in calibration.
		Right: 100ph/pixel/frame background (sub-aperture collects 40000ph/frame from the guide star).
		Input Tip-Tilt RMS is 255nm, measurement noise variance is 10nm${}^2$, integrator gain is 0.5 and
		total delay is 2 frames.}
		\label{fig:CoGbackGround}
	\end{center}
\end{figure*}
In this section we present two examples, showing the effect of a spot size change on
sensitivity for a quad-cell SH WFS and the effect of a background variation
on a Center of Gravity (CoG) SH WFS.
For both examples, we run a numerical simulation of a simple system, with a single
sub-aperture SH and with a phase containing only tip-tilt,
using the PASSATA software\cite{doi:10.1117/12.2233963}.\\
The parameters of the simulations are:
\begin{itemize}
\item WFS FoV = 2.5arcsec;
\item dimension of the sub-aperture (side) = 0.2m;
\item number of pixels of the detector (side) = 12;
\item WFS central wavelength = 750nm;
\item input Tip-Tilt RMS = 255nm.
\end{itemize}
In the first example we run the same simulation for two spot sizes:
in the first case with the Diffraction Limited (DL) spot (FWHM=0.77arcsec)
and the second case we convolve the DL spot with a 2D Gaussian shape with FWHM
of 8 pixels (FWHM=1.84arcsec).
In both cases we use the quad-cell algorithm to compute the slopes.
We know from calibration that the slope coefficient is 0.0080nm${}^{-1}$ for a DL spot and
0.0028nm${}^{-1}$ for a DL spot convolved with the 2D Gaussian shape with FWHM
of 8 pixels, hence, the relative sensitivity should be 0.35
(almost the ratio between the two spots size: 0.77/1.84).
The frequency range, $\left[\omega_1,\omega_2\right]$, chosen is $[50,500]$.
In Figure \ref{fig:quadCellSpot} the measurement PSD for the two cases are shown.
The estimated sensitivity loss factor of the first case is, as expected, 1 and
for the second case is about 0.35 (between the yellow and the green lines).\\
As for the first example in the second one we run the same simulation twice: the first time with
no background and the second time we add a background of 100ph/pixel/frame
(sub-aperture collects 40000ph/frame from the guide star).
In both cases we use the CoG algorithm to compute the slopes.
Sky background photons increase the denominator of the slope computation
of a factor 0.36 ($\frac{14400}{40000}$) and so the gain should be 0.74 ($\frac{1}{1+0.36}$).
The frequency range, $\left[\omega_1,\omega_2\right]$, chosen is again $[50,500]$.
In Figure \ref{fig:CoGbackGround} the measurement PSD for the two cases are shown.
The estimated sensitivity loss factor of the first case is, as expected, 1 and for second case
is about 0.75 (between light blue and blue-green lines).\\
With these two examples we saw how spot size and background noise affect the sensitivity
and how the sensitivity gain translates on modification of the TF, producing measurable effects on the PSD.
Actually, we estimate the sensitivity gain, $\alpha$, analysing the PSD.

\section{Conclusions}

An accurate estimate of the WFS sensitivity is essential to get the correct measurement
amplitude in the AO loop.
We show in this paper how the WFS sensitivity affects the transfer functions of
the closed loop, producing measurable effects on the closed loop and inputs data PSDs,
so that PSD analysis can be used to estimate the sensitivity gain.
The main limitation of this work is that, in most cases, a direct measurements
of the input data is not available.
This is the reason why further work is needed to find a method to estimate
the WFS sensitivity when only closed loop data is available.

On the other hand, a robustness check of the evaluation of the sensitivity gain with respect to the knowledge of the turbulence and with respect to WFS noise estimated, may provide information about the operative range of the method described herein.

\bibliographystyle{spiebib}       
\bibliography{biblio}   

\end{document}